\documentclass[12pt,fleqn]{article}
\usepackage{calc}
\usepackage{graphicx}
\usepackage{amssymb}
\usepackage{subfigure}
\usepackage{psfrag}
\usepackage{vmargin}

\newcommand{\xgobs}{x_\gamma^{obs}}
\newcommand{\xgll}{x_\gamma^{LL}}

\begin{document}


\setpapersize{A4}


\begin{titlepage}

\begin{center}\begin{bf}

\begin{LARGE}

NLL Monte-Carlo approach in 1 or 2 jets Photoproduction\footnote{This work was supported in part by the EU Fourth Framework Programme ``Training and Mobility of Researchers'', Network ``Quantum Chromodynamics and the Deep Structure of Elementary Particles'', contract FMRX-CT98-0194 (DG 12 - MIHT).} 

\end{LARGE}

\vskip 1.5cm

P. Aurenche$^a$, L. Bourhis$^c$, M. Fontannaz$^b$, J. Ph. Guillet$^a$

\end{bf}\end{center}

\vskip 0.5cm

\hskip 2 truecm$^a$ Laboratoire de Physique Th\'eorique LAPTH\footnote{URA 14-36 du CNRS, associ\'ee \`a l'Universit\'e de Savoie.},
\par \hskip 2 truecm B.P. 110, F-74941 Annecy-le-Vieux Cedex, France\\

\hskip 2 truecm $^b$ Laboratoire de Physique Th\'eorique\footnote{Unit\'e Mixte de Recherche CNRS - UMR 8627},
\par \hskip 2 truecm Universit\'e de Paris XI, b\^atiment 210, F-91405 Orsay Cedex, France\\

\hskip 2 truecm $^c$ Department of Theoretical Physics, \par
\hskip 2 truecm South Road, Durham City DH1 3LE, United Kingdom \\

\vskip 2cm

\begin{abstract}
We present a method for calculating the photoproduction of jets at HERA based on
Next to Leading Logarithm QCD calculations. It is implemented in  a Monte-Carlo generator which
allows us to easily compute any infra-red  safe cross sections for 1 or 2 jet observables using
various jet reconstruction algorithms. We focus on the possibility of extracting the gluon
contents of the photon and of the proton from present and future H1 and ZEUS data. 
\end{abstract}

\vskip 2 truecm
\hfill Durham 00/26 \par
\hfill LAPTH-785/00 \par
\hfill LPT-Orsay/99/95 \par
\hfill hep-ph/0006011 

\end{titlepage}

\section{Introduction}
\label{Introduction}

Electron-proton scattering at HERA is dominated by the exchange of a quasi-real
photon and a fraction of $\gamma$-$p$ collisions leads to the production of
high $p_T$ jets. Therefore these processes can be predicted by perturbative QCD and they  can be used to investigate the
structure of the photon in a way complementary to the study of the deep
inelastic scattering $e$-$\gamma$. In the  latter case, one can measure directly
the quark density inside the photon whereas the gluon density is constrained
by the evolution equation.  On the contrary, in photoproduction, one can probe
directly the gluon contents of the photon. 

Leading Logarithm (LL) results have been available for a long time, but they are plagued
by sizeable uncertainties coming from the dependence on unphysical  scales.
Therefore, Next to Leading Logarithm (NLL) calculations are essential to describe the
photoproduction of jets. To reach this goal, we have adapted a  method
developped previously to deal with the production of two high-$p_T$  hadrons in
hadron collisions \cite{CFG}. We apply it to build a Monte-Carlo generator which
is able to produce a set of partonic events on which  one can apply any jet
reconstruction algorithm to produce a set of jet  events. With the latter we can
easily compute any infrared safe cross sections. This technique gives a lot of
flexibility to the study of various  jet algorithms and observables.

In particular we will apply it to the study of two jet observables in which the
distribution function variables are well constrained by kinematics. This allows
us to disentangle the distribution functions from the hard subprocess more
easily, because less convolutions are involved in the calculation of the cross
sections. Some results obtained with this approach have already been reported in
ref.~\cite{2r}.
 
The paper is organised as follows. The theoretical framework is described in
section~\ref{theory}. We also compare our approach with those of other authors. In
section~\ref{1jet} we present some applications of our work to 1-jet cross sections and we
compare our results with those already obtained with an analytical approach.
Section~\ref{2jet} is devoted to the study of 2-jet observables. We will examine whether it is
possible to accurately measure the distribution functions of the gluon in the photon and in the
proton. In section~\ref{h1-zeus} we study recent H1 and ZEUS data and the constraints they put
on the gluon distributions. The conclusions are in section~\ref{conclusion}.

\section{Description of the method}
\label{theory}

To calculate cross sections and isolate collinear and infrared singularities, we
used a ``phase space slicing method''. We start from the $2 \rightarrow  3$
partonic squared matrix elements and virtual corrections evaluated by  Ellis and
Sexton \cite{ES}. Collinear and infra-red singularities lead to  poles in
$1/\epsilon$ and $1/\epsilon^2$ when using dimensional regularisation. For a
generic real process $1 + 2 \rightarrow 3 + 4 + 5$, at least two partons have high
transverse energy $E_T$ (3 and 4) and only one can be soft (5). In order to 
extract the singularities, we cut the phase space in several parts: part I where
$E_{T 5}$ is less than a given scale $p_{Tm}$ and part II where $E_{T 5}
>p_{Tm}$. Part II is divided in three parts: IIa (resp. IIb) where parton 5 is
within a cone around parton 3 (resp. 4) called $C_3$  (resp. $C_4$), IIc where
parton 5 is outside $C_3$ and $C_4$. Parton 5 is in $C_i$ if
$((\phi_5-\phi_i)^2+(\eta_5-\eta_i)^2)^\frac{1}{2} <  R_c$. Here $\eta =
- \log
\tan(\theta/2)$ is the pseudo-rapidity and $\phi$  is the azimuthal angle. Part
I contains infra-red singularities and collinear singularities in the initial
state and parts IIa and IIb contain collinear singularities in the final state
whereas part IIc is finite. 

The contributions of regions I, IIa and IIb are calculated analytically and the infra-red singularities are cancelled by the corresponding ones in the virtual terms. Initial collinear singularities are factorised in the  parton distributions and the final collinear singularities disappear when  integrating on the relative momentum between parton 5 and the parton  with which it is collinear due to energy momentum sum rules.

The finite parts remaining after the cancellation of singularities have
been analytically computed using Maple \cite{Maple}. Large logarithms
$\log p_{Tm}$, $\log^2 p_{Tm}$ and $\log R_c$ appear in the collinear  and infrared contributions. They are cancelled by similar terms from  part IIc so that the total cross section is independent of these unphysical cuts. It should be noted that we have kept only the logarithmic terms in the calculation of contributions I, IIa and IIb, neglecting terms of order ${\cal O}(p_{Tm} \log p_{Tm})$, ${\cal O}(R_c^2 \log p_{Tm})$, and less singular terms.

Using the Monte-Carlo package BASES \cite{bases}, our program generates quasi $2
\rightarrow 2$ events corresponding to collinear, Born and virtual contributions (we
take $p_{Tm} \ll E_{T_{3,4}}^{min}$ and $R_c \ll  1$) and $2 \rightarrow 3$ events
corresponding to part IIc. For the latter, a jet reconstruction algorithm is applied.
Finally, these events are histogrammed~\cite{hbook} in order to give any cross
sections we are interested in. We have checked that the cross sections do not depend
on the unphysical cuts $p_{Tm}$ and $R_c$ in the region where $0.005<p_{Tm}<0.1$ and
$0.01<R _c<0.1$. From a numerical point of view, the compensation between this cut
dependence arises mainly between positive real contribution of part IIc and negative
contributions of part I, IIa, IIb and virtual corrections~; we need to generate a
sufficient number of events in  order to obtain a small error after the compensation.
We take the greatest  possible values for the cuts $p_{Tm}$ and $R_c$ (i.e. $p_{Tm}= 0.1\ GeV$  and $R_c= 0.1$) in order to lower the size of this
compensation which is  then typically of the order of 1 for 5.

This approach is applied to the direct and resolved (proportional to the photon distribution functions) parts of the cross section. However these contributions depend on the convention adopted in the subtraction of the collinear singularities. Unless explicitly specified, we use the $\overline{MS}$ factorization and renormalization schemes.

Several authors have used similar approaches in the calculation of
jet-pho\-to\-pro\-duc\-tion cross sections. Harris and Owens \cite{7bisr} also built an
event generator~; but their phase slicing is based on cuts put on $E_5$ (in the parton CM)
and on the Mandelstam variables of the subprocess which control the collinear divergences.
Klasen and Kramer \cite{8bisr} introduced a single invariant cut on the Mandelstam variables. On the other
hand Frixione and Ridolfi \cite{9bisr} made their calculations with the subtraction method
\cite{10bisr}. We made several numerical comparisons with the results of Klasen and Kramer,
and found a good agreement.


Finally let us note that it is very convenient to use variables $E_{T}$, $\eta$ and $\phi$
to define regions of the phase space in which we perform
analytical calculations. Indeed these are the variables commonly used by
experimentalists to define jets and cuts in their phase space. If $p_{T_m}$ and $R_c$ are smaller than respectively the lower bound of the transverse momenta and the width of jets, we never
integrate (thus defining an inclusive measurement) in a phase space region in
which the experimentalists perform an exclusive measurement. Since these experimental bounds are much larger than the domain in which our neglecting non logarithmic terms is valid this is in fact not a constraint at all. On the contrary with the invariant cut method, this can only be achieved with very small cuts.

\section{Single jets}
\label{1jet}

In this section we present some numerical results  for one-jet cross sections in order to compare our predictions with those obtained in a preceding paper \cite{AFG-1jet}. Our inputs are the following. At the HERA collider ($\sqrt{s} = 300$~GeV), electrons produce photons with small virtuality $Q^2$. We use the kinematical conditions of the ZEUS  collaboration \cite{8r}: $Q^2_{max} = 1\ GeV^2$ and $0.2 < y < 0.85$ where $y=E_\gamma /E_e$. The spectrum of the quasi-real photon is approximated by the  Weizs\"acker-Williams formula. 

\begin{eqnarray}
F_e^{\gamma}(y) = {\alpha \over 2 \pi} \left \{ {1 + (1 - y)^2 \over y} \ \log \ {Q_{max}^2(1 - y ) \over m_e^2 \ y^2} - {2(1 - y) \over y} \right \} \quad .
\label{1e}
\end{eqnarray}

For the proton distributions we take the CTEQ4M parametrization \cite{9r} and for the photon
distributions the GRV parametrization \cite{10r} transformed to the $\overline{MS}$ scheme. We
use five flavours, $\Lambda_{\overline{MS}}^{(4)} = 296$~MeV, a renormalization scale $\mu$ and
a factorization scale $M$ equal to the transverse energy $E_{T}$ of the jet. For all our
calculations, we use for $\alpha_s(\mu )$ an exact solution of the two-loop renormalization group
equation, and not an expansion in $\log {\mu \over \Lambda}$. The cross sections are higher by
some 2.5~\% when the exact $\alpha_s (\mu )$ is used with $\mu \sim 15$~GeV.

\begin{figure} \centering
\begin{minipage}{\textwidth} \centering \includegraphics[width=0.9\textwidth]{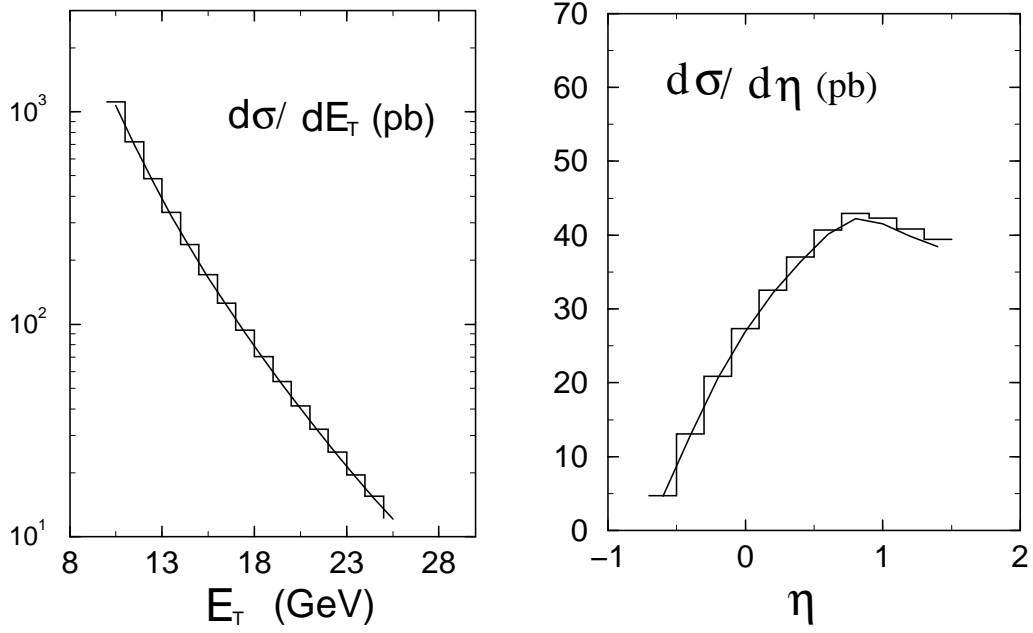}
\end{minipage} \caption{a) Single jet resolved cross section $d\sigma/dE_T$ integrated in the
rapidity range $.5 \leq \eta \leq 1.5$ (histogram) and compared to the analytical result of
\protect{\cite{AFG-1jet}}. b) $d\sigma/d\eta$ for the resolved contribution with 20 GeV $<
E_{T} < 21$~GeV. } \label{fig:1jet} \end{figure}

Here we present results only for the resolved contribution. Jets are defined with the
$k_T$-algorithm \cite{11r}. We compare our results with those of a previous analytical
calculation~\cite{AFG-1jet}, for the transverse momentum distribution in Fig.~\ref{fig:1jet}a
and for the rapidity distribution in Fig.~\ref{fig:1jet}b. We can see that the agreement is
quite good between these two sets of results. Similar comparisons hold for the direct part. 

We do not pursue the study of 1-jet cross sections, because they do not constrain the parton distributions as well as the 2-jet cross sections do~; the latter offer more kinematical possibilities of control of the $x$-variable of these distributions.   

\section{Dijets cross sections and the photon structure function}
\label{2jet}

In this section we study the dijet cross sections and put the emphasis on variables and cross sections which give access to the gluon distribution functions. Comparison between data and theory is postponed to section 5. The dijet cross section, as  a function of the transverse energy $E_{T 3}$ and the jet rapidities $\eta_3$ and $\eta_4$, is given by a product if the subprocess is a $2 \to 2$ reaction (LL approximation) 

\begin{eqnarray}
{d\sigma \over dE_{T 3}^2
d\eta_3 d\eta_4} = \sum_{a,b,c,d} x_e \ F_e^a(x_{e},M)\  x_p \ F_p^b (x_{p},M) {d\sigma_{ab \to cd}
(\mu) \over dt} \ \ ,\label{LL-pt-etas}
\end{eqnarray}

\noindent 
$d\sigma/dt$ is the $a + b \to c + d$ cross section, and $F_e^a(F_p^b )$ the parton distributions in the electron (proton) ($x_p = E_{T 3}(e^{\eta_3} + e^{\eta_4})/2 E_p$ and $x_e = E_{T 3}(e^{-\eta_3} + e^{-\eta_4})/2 E_e$)~; $M$ is the factorization scale and $\mu$ the renormalization scale. The direct contribution corresponds to $a = \gamma$ and $F_e^{\gamma}$ is the Weizs\"acker-Williams formula (\ref{1e}). In the resolved case, $F_e^a$ is given by a convolution of $F_e^{\gamma}$ with the parton distribution in the photon 

\begin{eqnarray}
F_e^a(x_e,M) = \int_0^1 dy \ dx_{\gamma} \ F_e^{\gamma}(y) \
F_{\gamma}^a(x_{\gamma},M)\  \delta (x_{\gamma}y - x_e) \ \ . \label{Fe}
\end{eqnarray}

\noindent  If the photon energy $E_{\gamma}= yE_e$ is known (for
instance by tagging the outgoing electron), we can measure $x_{\gamma}$
and define the observable

\begin{eqnarray}
{d\sigma \over dx_{\gamma}} &=& \sum_{a,b,c,d} x_{\gamma} \
F_{\gamma}^a(x_{\gamma},M) \int_0^1 dy \ dE_{T 3}^2  d\eta_3 d\eta_4 \
\delta \left ( x_{\gamma} - {E_{T 3} \left (e^{-\eta 3} + e^{- \eta_4}
\right ) \over E_{\gamma}} \right )\nonumber\\
&&\times \ F_e^{\gamma}(y)\  x_p \ F_p^b (x_p,M) {d\sigma \over dt} \ .
\label{LL-xgamobs}
\end{eqnarray}

\noindent Thus the dijet cross section written as a function of $x_{\gamma}$ is
proportional to the parton distribution in the photon $F_{\gamma}^a(x_{\gamma})$.
One observes that the direct contribution, with $F_{\gamma}^a = \delta_{\gamma_a} \delta (1 -
x_{\gamma})$, leads to a peak in the cross section at $x_{\gamma}=1$.
 
 \begin{figure}[h]
	\centering
	\begin{minipage}{0.7\textwidth}
		\centering
		\includegraphics[width=0.9\textwidth]{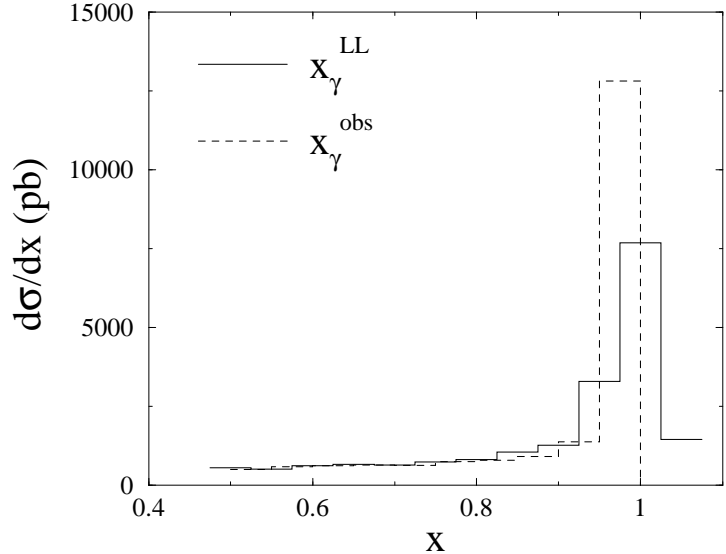}
\end{minipage}%
\caption{A comparison between the distributions $d\sigma/d\xgll$ and $d\sigma/d\xgobs$, with the following cuts on the two high
$E_T$ jets: $E_T^{leading} > 12\ GeV$ and $E_T^{trailing} > 10\ GeV$ and $0<\eta<1$,
and the $k_T$-algorithm for the definition of jets. }
\label{fig:xgamLL-xgamOBS} \end{figure}

NLL QCD corrections to the LL expression (\ref{LL-xgamobs}) blur
its simple kinematics. There are contributions with three jets in  the final
state, and we can no longer fix $x_{\gamma}= \sum\limits_{i=3}^5 {E_{T i} e^{-\eta_i} \over 2
E_{\gamma}}$, because the third jet is not observed. However we can follow the
strategy of the ZEUS collaboration \cite{zeus} which defines the variable (jets 3
and 4 are the jets with the highest transverse energies)
\begin{eqnarray}
\xgobs = \frac{\left ( E_{T 3} e^{- \eta_3} + E_{T 4} e^{- \eta_4} \right )}{2E_{\gamma}}
\label{def-xgamobs}
\end{eqnarray}
\noindent and observe the dijet cross section $d\sigma/d\xgobs$. However this
definition of $\xgobs$ may lead to infrared sensitive cross sections. Indeed
fixing $E_{T 3}$ \underbar{and} $E_{T 4}$ strongly constrains the available
phase space of the unobserved partons (for instance parton 5 in section 2) and
forbids a complete compensation between real and virtual NLL corrections. This
results in cross sections containing, for instance, terms proportional to $\log \left ( 1 - \left ( E_{T 4}^{min} / E_{T 3}\right )^2 \right )$ after
an integration over $E_{T 4}$ from a lower bound $E_{T 4}^{min}$ (smaller
than $E_{T 3}$) has been performed. The cross section is not defined at
$E_{T 3} = E_{T 4}^{min}$, although it is integrable. Therefore the variable
$x_{\gamma}^{obs}$ defined in (\ref{def-xgamobs}) should only be used with cross
sections integrated over $E_{T 3}$ and $E_{T 4}$~; moreover the integration
range of $E_{T 3}$ and $E_{T 4}$ should not have the same bounds. A
discussion of this condition may be found in ref. \cite{9bisr}. 

For these reasons, we prefer the variable
\begin{eqnarray}
\xgll = \frac{E_{T 3} \left ( e^{-\eta_3} + e^{-\eta_4} \right )}{2E_{\gamma}}
\label{xgamLL}
\end{eqnarray}
\noindent which depends on the transverse energy of only one jet. It can be
associated with cross sections in which the energy of the second jet is not
measured. One observes that $x_{\gamma}^{LL}$ may take values larger than 1.0. 

One must also observe that the cross sections $d\sigma/dx_{\gamma}^{LL}$ or $d\sigma/dx_{\gamma}^{obs}$ are singular when $x_{\gamma}^{LL}$ or $x_{\gamma}^{obs}$ approaches 1. Indeed we have the condition

$$1 \geq {E_{T 3} \ e^{-\eta_3} + E_{T 4} \ e^{- \eta 4} + E_{T 5} \ e^{-\eta_5} \over 2 E_{\gamma}}$$

\noindent (the sign = corresponds to the direct case), which means

\begin{eqnarray}
1 - x_{\gamma}^{obs} \geq {E_{T 5} \ e^{-\eta_5} \over 2 E_{\gamma}} \quad .
\label{7e}
\end{eqnarray}

\noindent When $x_{\gamma}^{obs}$ goes to 1, the phase space of parton 5
is severely restricted. This results in $\log (1 - x_{\gamma}^{obs})$
terms generated by the NLL corrections. We obtain a similar result with
$x_{\gamma}^{LL} \to 1$, although the phase space of parton 5 is less
severely constrained than by condition (\ref{7e}). Therefore we expect a
smoother behaviour of $d\sigma /dx_{\gamma}^{LL}$. This point can be
verified in Fig.~\ref{fig:xgamLL-xgamOBS} in which we display $d\sigma
/dx_{\gamma}^{obs}$ and $d\sigma/dx_{\gamma}^{LL}$ for the direct term. 
Besides the region very close to 1.0, the two distributions are very
similar \footnote{Note that the shape of the cross
sections around $x_{\gamma} = 1$ depends on the width of the $x_{\gamma}$-bin, because
the Born and virtual contributions are proportional to $\delta (1 - x_{\gamma})$. If
the width is too small, the cross section may even be negative (the positive real
contributions do not compensate anymore the negative virtual contributions).}.  (The AFG \cite{13r} and ABFOW \cite{18r} distributions with $N_f = 4$  have been used for this calculation). Currently experimentalists integrate $x_{\gamma}^{obs}$ over the range $.75 \leq x_{\gamma}^{obs} \leq 1$. Therefore the singular behaviour of the cross section is smoothed and should not forbid the phenonemological application of NLL calculations.

\begin{figure}
	\psfrag{ms}[lc][lc][0.8][0]{$\overline{MS}$}
	\psfrag{disg}[lc][lc][0.8][0]{$DIS_\gamma$}
	\psfrag{AFG}[lc][lc][0.8][0]{AFG}
	\psfrag{GRV}[lc][lc][0.8][0]{GRV}
	\psfrag{xgll}{$\xgll$}
	\psfrag{dsig}{$d\sigma/d\xgll$ (pb)}
	\centering
	\begin{minipage}[b]{0.45\textwidth}
	\centering
	\subfigure[Direct contribution]{
		\includegraphics[width=\textwidth]{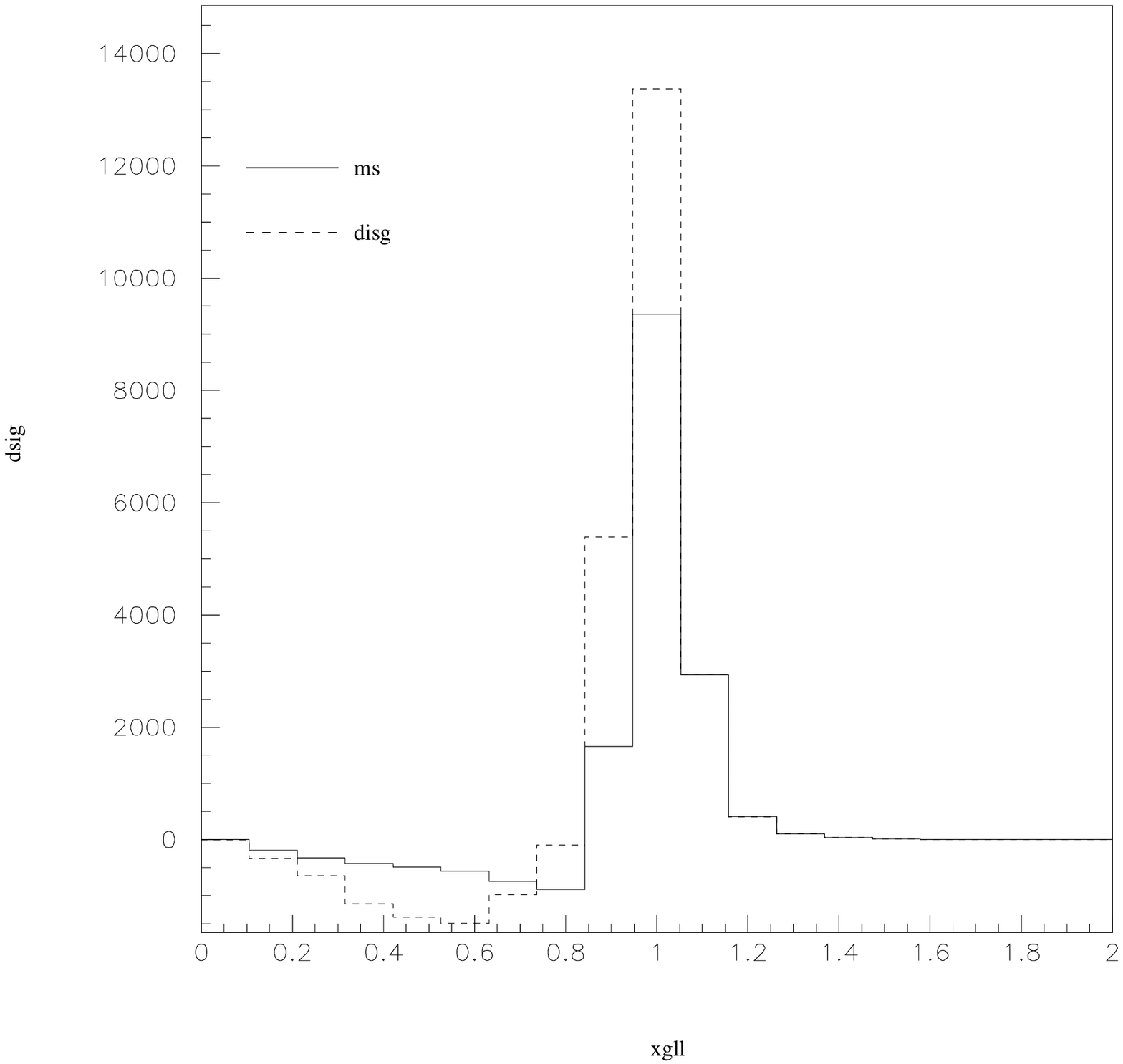}
	}%
	\end{minipage}%
	\begin{minipage}[b]{0.45\textwidth}
	\centering
	\subfigure[Resolved contribution]{
		\includegraphics[width=\textwidth]{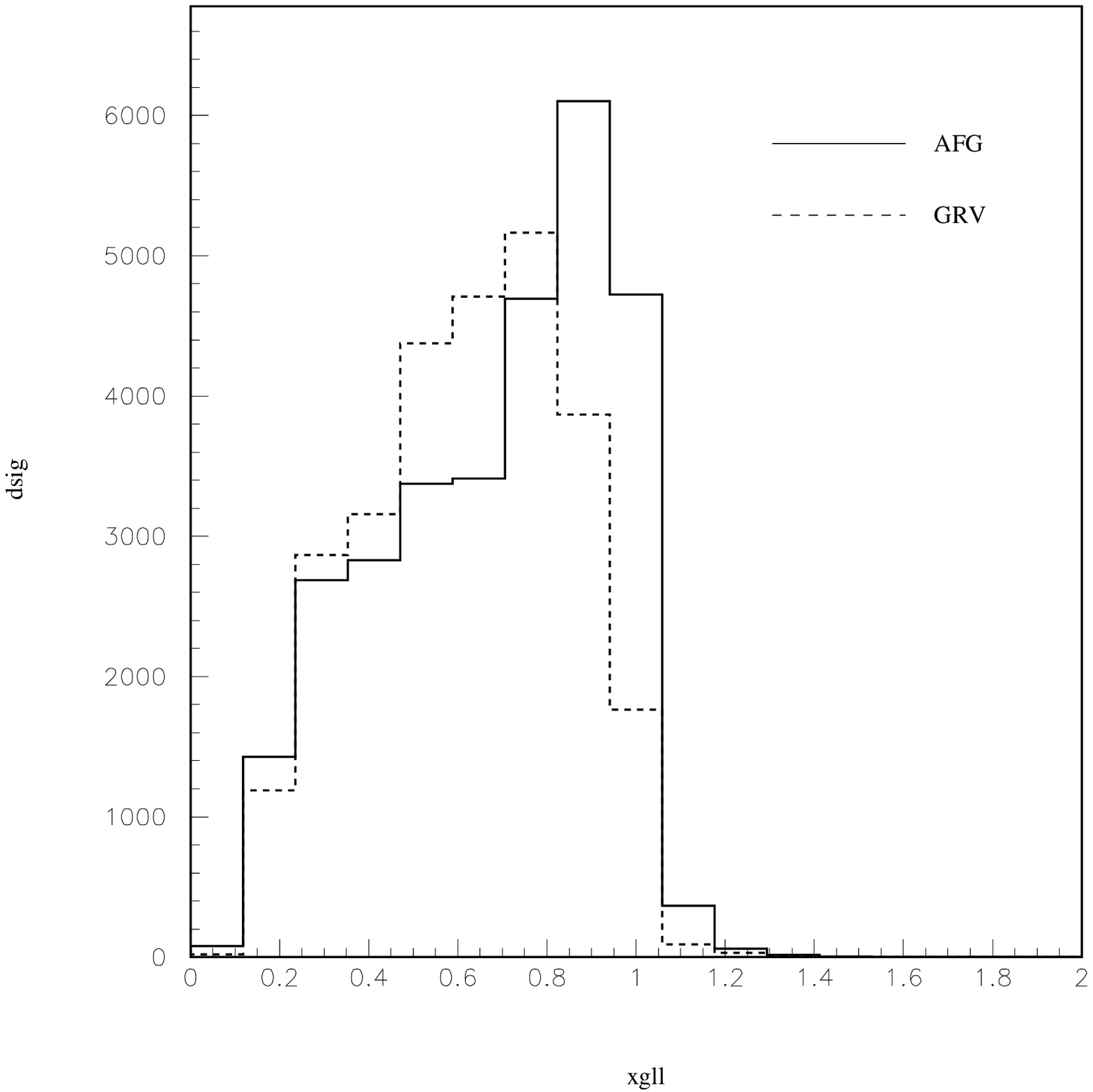}
	}%
	\end{minipage}
	\begin{minipage}[b]{0.50\textwidth}
	\centering
	\subfigure[Total contribution]{
		\includegraphics[width=\textwidth]{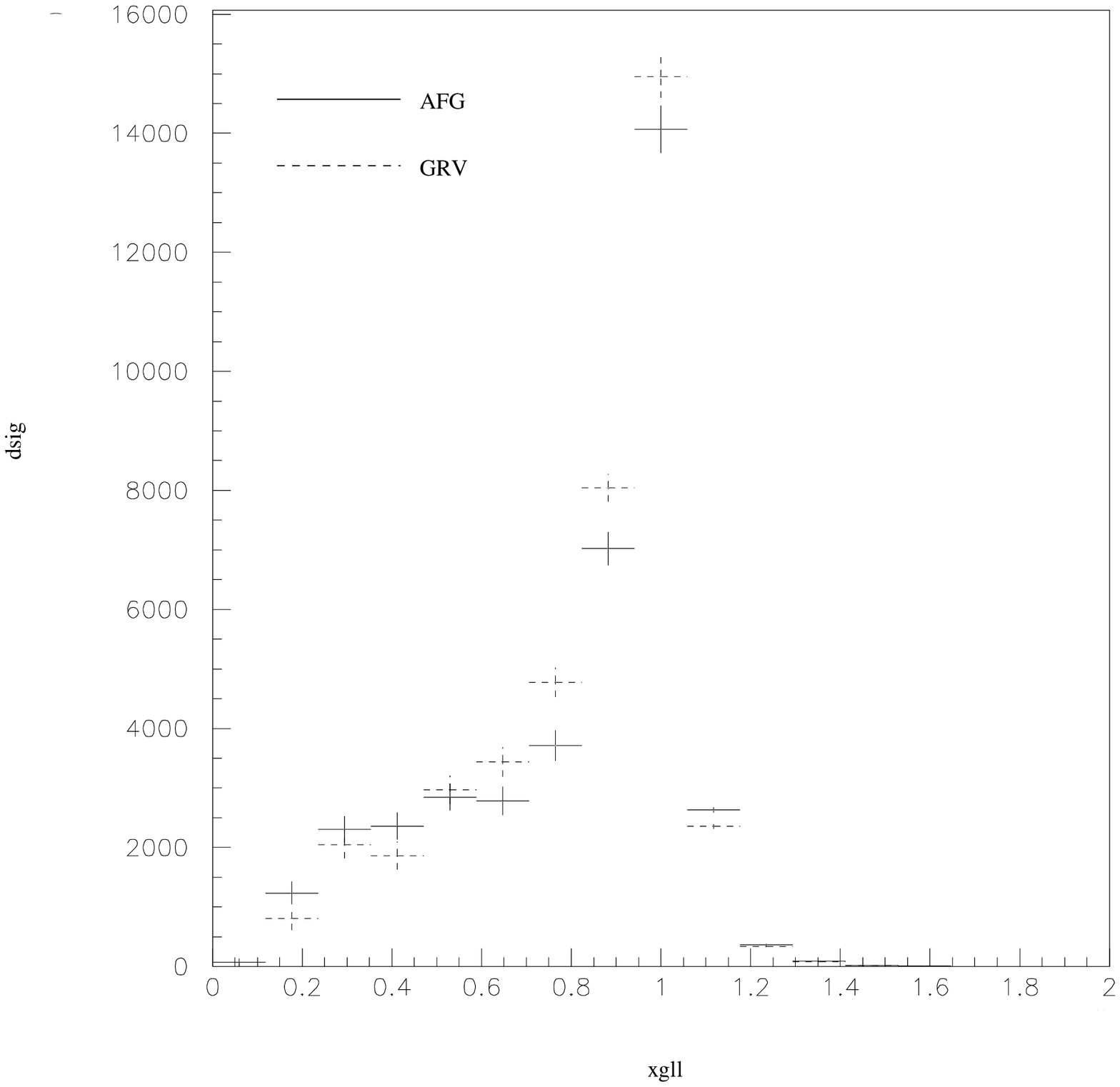}
	}%
	\end{minipage}%
\caption{Distribution $d\sigma/d\xgll$ for the direct, resolved and total contributions with the AFG ($\overline{MS}$ scheme) and GRV (DIS$_\gamma$ scheme) photon
density as well as ABFOW proton density, with the following cuts on the two high
$E_T$ jets: $E_T > 9.2\ GeV$ and $-1<\eta<2$, and using a cone algorithm with
$R=1$ and no $R_{sep}$ to define the jets.} \label{fig:xgamLL-afg-grv} \end{figure}

We can also see from Fig.~\ref{fig:xgamLL-xgamOBS} how much NLL corrections modify the $x_{\gamma}$ distribution which is proportional to $\delta (1 - x_{\gamma})$ at the LL accuracy. NLL corrections generate terms with $x_{\gamma}^{obs}$ or $x_{\gamma}^{LL}$  different from 1 and the simple picture of a photon directly interacting with a quark of the hard subprocess at $x_{\gamma} = 1$ is lost. But one must also keep in mind that the separation of the cross section $d\sigma /dx_{\gamma}^{obs}$ into a direct part and a resolved part is factorization scheme dependent and that $d\sigma^{direct}/dx_{\gamma}^{obs}$ has no physical sense on its own (the same remark is valid for $x_{\gamma}^{LL}$).  

To study this problem we computed the observable $d\sigma/dx_{\gamma}^{LL}$ in two
factorization schemes, the so-called $DIS_{\gamma}$ and $\overline{MS}$ ones. For  the resolved
part, we used for the former case the GRV distributions of  quarks inside the photon and for
the latter the AFG distributions. (Here we also use the ABFOW proton distributions). We can  see that both the direct and resolved part are very
different in these two schemes (Fig.~\ref{fig:xgamLL-afg-grv}), but that this difference is
much smaller for the total cross section which is factorization scheme independent. The
remaining difference partly comes from different hadronic inputs in  the two sets of
distributions. 

\begin{figure}
\centering
\begin{minipage}{0.5\textwidth}
\subfigure[$0 < \eta < 1$]{
\includegraphics[angle=-90,width=0.9\textwidth]{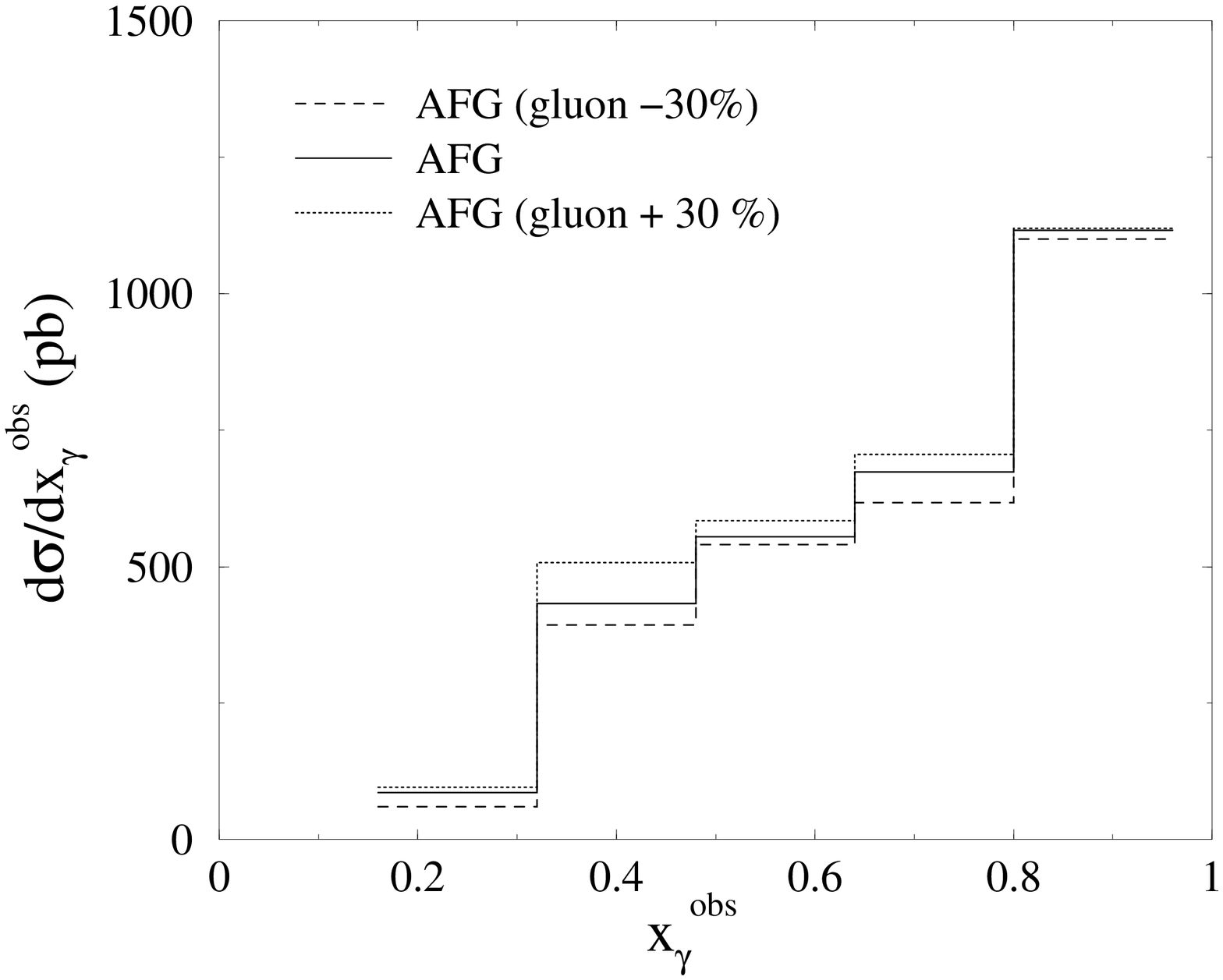}
}%
\end{minipage}%
\begin{minipage}{0.5\textwidth}
\subfigure[$1 < \eta < 2$]{
\includegraphics[angle=-90,width=0.9\textwidth]{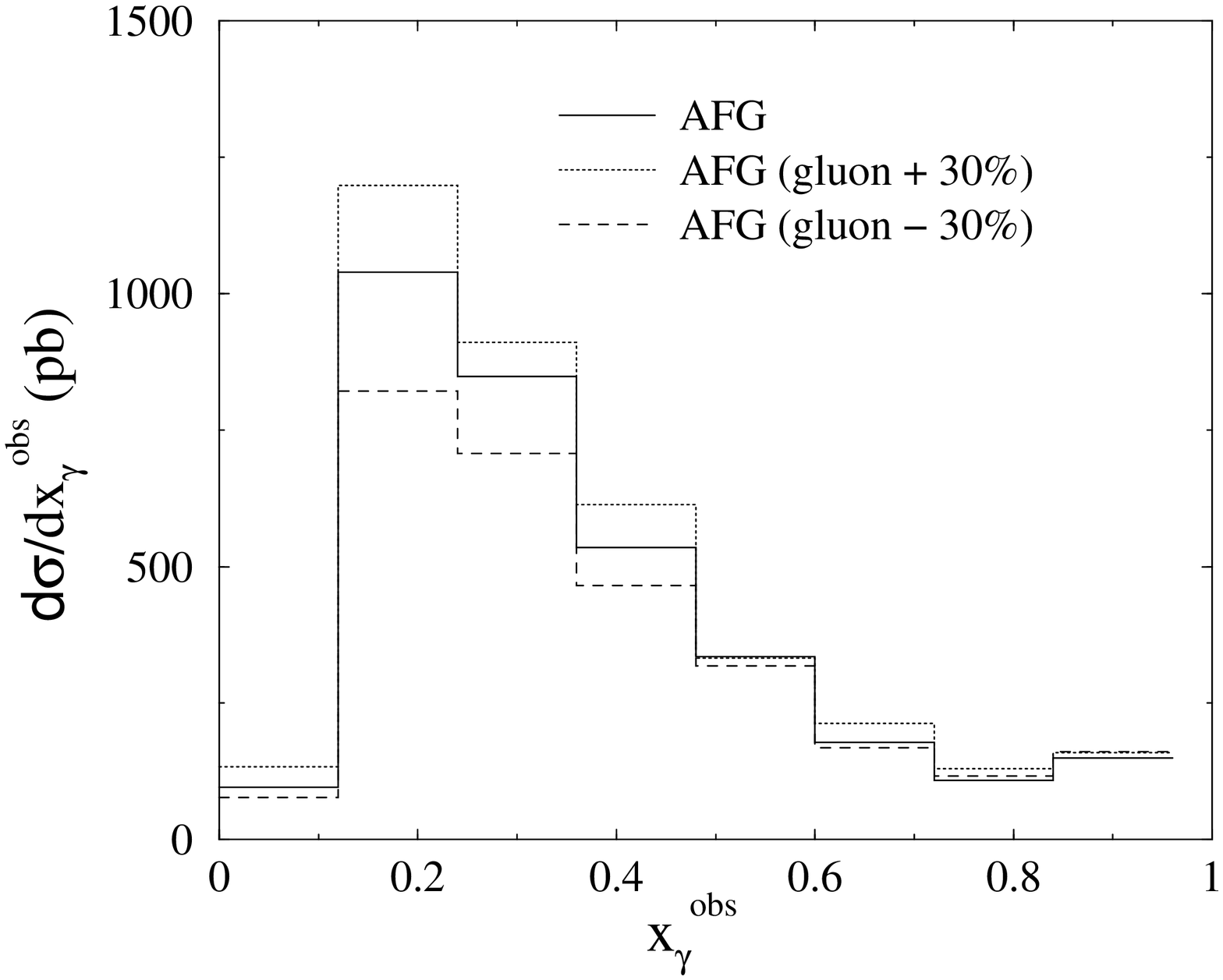}
}
\end{minipage}
\caption{Cross sections $d\sigma/d\xgobs$ with the following cuts for
the two high $E_T$ jets: $E_T^{leading} > 12\ GeV$ and $E_T^{trailing} >
10\ GeV$ and 2 different cuts on their rapidities, and using the
$k_T$-algorithm to define the jets. Curves with the gluon distribution
varied by $\pm 30\%$ around the AFG one are also displayed.}
\label{fig:xgamobs-varglue} 
\end{figure}

We now turn to the study of the gluon contents of the photon and to the
possibility of constraining it through the observation of the distribution
$d\sigma/dx_{\gamma}^{obs}$. The sensitivity of $d\sigma/dx_{\gamma}^{obs}$ to the
gluon distribution does not only depend on the value of $x_{\gamma}^{obs}$, but
also on the various kinematical cuts imposed on the 2-jet phase space. Clearly the
region of positive and large rapidities corresponds to small values of
$x_{\gamma}^{obs}$ and to an enhancement of $d\sigma/dx_{\gamma}^{obs}$. This is
verified in fig. 4. We divided the rapidity range in the lab frame for the two
leading jets in  
subintervals: 
$0 < \eta < 1$ and $1 < \eta < 2$. We have also used asymmetric cuts on the transverse momenta, as it is done by experimental collaborations, in order to avoid the instabilities which appear when these momenta become close to each other. Then in each of these rapidity intervals we compared the $\xgobs$ distribution obtained with the AFG photon
density with cross-sections for which we have artificially reduced and increased
the gluon distribution by 30~\%. We found that the influence of the
gluon increases with the rapidities and in fact the differences between these
three curves become sizeable only when $1 < \eta < 2$. Indeed an increase of 30~\%
of the gluon density results in an increase of approximately 25~\% for the
cross-section around $\xgobs = 0.2$. Therefore a determination of the gluon
contents of the photon would require to use such cuts on rapidities. Such a study
would be able to test the gluon density $F^{g}_{\gamma}(x)$ in the region $x \approx
0.2$ where it is very poorly known.

\section{Comparison with H1 and ZEUS data}
\label{h1-zeus}

In this section we analyse recent H1 and ZEUS data to assess the possibility to
put constraints on the gluon distributions in the photon and in the
proton. With this intention we investigate the sensitivity of various
cross sections to changes in the gluon distributions, and we compare
their variations to the experimental errors. In this way we obtained
an estimate of the accuracy with which gluon distributions can be 
extracted from present data, and from future high statistics
experiments.
We modify the gluon distributions by increasing or decreasing their
normalizations by a few tens of percents. This method has the advantage
of leaving unchanged the well-determined quark distributions and to
quantify the gluon modifications in a simple way. Of course gluon
distributions are constrained by other experiments. But, as discussed in
the introduction, the gluon in the photon is not well constrained by DIS
$\gamma\-\gamma^*$ data. As for the gluon in the proton we are going to study observables sensitive to the distribution at small $x \approx 0.02$. In this region the gluon is pinpointed by the slope of $F_2$ and it was shown by the CTEQ collaboration that a variation of about $\pm 10\%$ of $F^g_p$ in this region and at a scale of 100 GeV would cause clear disagreements with DIS plus Drell-Yan data \cite{cteq98}. However this conclusion was not obtained by performing an error analysis but only by tuning the gluon parametrisation, which might artificially reduce the range of variation by being too restrictive. A more recent study obtained a gluon density 30\% bigger than CTEQ4M about $x \approx 0.02$ and at a scale of 20 GeV \cite{barone}. Errors on the gluon determination are also presented in ref. \cite{19ref}. Thus we think that it is interesting to find how much photoproduction can constrain this density in this x-range.


The interested reader may find more global comparisons
between theory and data in ref \cite{14r,21r,16r}. The inputs we use in this
section have been defined in section 3 where we computed single-jet
cross sections. More details on the kinematical parameters used for dijet cross sections are given
below\footnote{To follow H1 and ZEUS conventions, we call jet 1 and jet 2 the jets
with the highest energies~; we used the labels 3 and 4 in the preceeding sections.}.


\begin{figure}
\centering
\begin{minipage}{0.7\textwidth}
\centering
\includegraphics[width=0.9\textwidth]{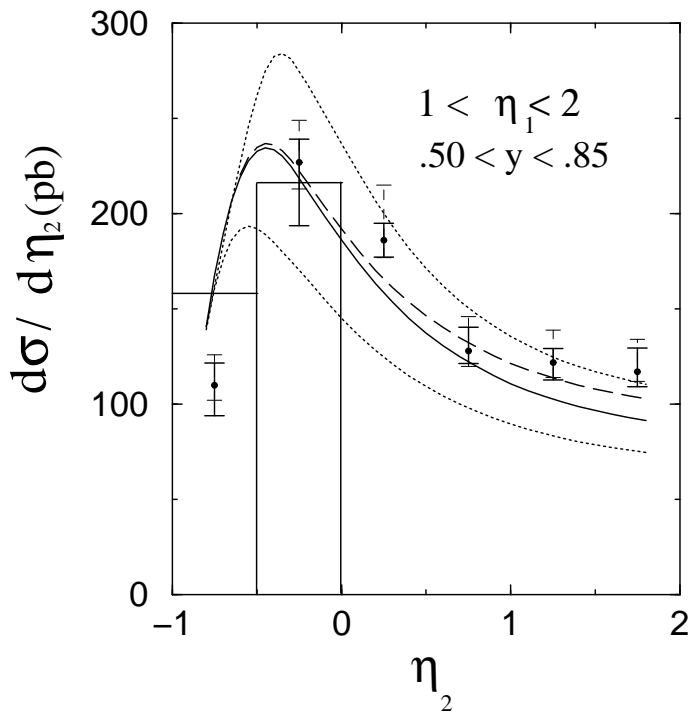}
\end{minipage}
\caption{The cross section $d\sigma/d\eta_2$ compared with ZEUS data \cite{16r}. (Full error
bars~: statistical errors~; dashed error bars~: systematic errors including energy-calibration
errors).  Full line: $.5 \leq y \leq .85$. Dotted lines: $.45 \leq y \leq .85$ (upper) and $.55
\leq y \leq .85$ (lower). The dashed line is obtained with the distribution $g/\gamma$
increased by 20~\%. } \label{fig:5} \end{figure}

The ZEUS collaboration presents cross sections which depend on the jet transverse momenta,
or on the jet rapidities \cite{16r}. Therefore we do not have a direct access to $d\sigma
/dx_{\gamma}^{obs}$, however we can study data corresponding to regions of the jet phase
space in which the role of the gluon in the photon or that of the gluon in the proton is
enhanced. These regions are essentially defined by large or small values of the jet
rapidities: large rapidities correspond to small $x_{\gamma}$ and small rapidities to small
$x_{p}$ (see the LL expressions for $x_p$ and $x_e$ below formula (\ref{LL-pt-etas})). 

Let us start with the case of large jet rapidities. Experiment and theory are compared in
Fig.~\ref{fig:5}.  The dijet cross sections in this figure correspond to events with at
least one jet with transverse energy larger than 14 GeV, the transverse energy of the other
jet being larger than 11 GeV~; it is integrated over $1  < \eta_1 \leq 2$. The real photon
kinematical domain is specified by $Q_{max}^2 = 1 \ {\rm GeV}^2$ and $.5 < y < .85$ (cf
expression (\ref{1e})). The jets are defined with the $k_T$-algorithm \cite{11r} and the
scales $M$ and $\mu$ are set equal to the transverse energy of the most energetic jet.

To test the sensitivity of $d\sigma /d\eta_2$ to the gluon density in the photon,
we increased the gluon distribution uniformly by a factor of 1.2. Comparing the
full and dashed curves, we see that this factor produces an increase of the cross
section for large values of $\eta_2$ by some 10~\%. As expected the backward region,
corresponding to large values of $x_{\gamma}$, is not affected~; the gluon distribution in
the photon decreases much faster than the quark distributions. In this region $d\sigma /d\eta_2$ rapidly varies with
$\eta_2$ and better comparison with data is obtained by integrating $d\sigma
/d\eta_2$ in the experimental bins. This has been done for the two bins: $-1. \leq
\eta_2 \leq - .5$ and $-.5 \leq   \eta_2 \leq .0$. The agreement between theory
and data is quite good in the second bin but the data at larger $\eta_2$ favor a larger gluon distribution in the proton. But one must notice that the present experimental
errors are larger than the effect produced by an increase of the gluon
distribution by 20~\%. In the backward region, a clear discrepancy appears which
could be attributed to hadronization effects as discussed in ref. \cite{23ref}.

It is also worth again noting the great sensitivity of the cross section to the $y$-range of
the photon. The energy of the incoming photon is reconstructed from the final hadron energies
with the Jacquet-Blondel method. Various corrections have to be applied to this ``photon
energy'' $y_{_{JB}}$ in order to obtain the true photon energy $y$ \cite{17r}. In
Fig.~\ref{fig:5} we show the effect of a 10~\% error on the determination of the lower bound of
the $y$ variable. It is very large for negative values of $\eta_2$. Even in the forward region,
this 10~\% error produces an effect much larger than the 20~\% variation of the gluon
distribution. Actually this ``error'' is included in the discussion of the systematic errors quoted in
\cite{17r}. But because the dispersion of $y_{_{JB}}$ around $y$ may reach 10~\%, a better
theoretical prediction could be obtained by taking into account the dispersion of the upper
and lower bounds of the variable $y$.

\begin{figure}
\centering
\begin{minipage}{0.8\textwidth}
\centering
\includegraphics[width=0.9\textwidth]{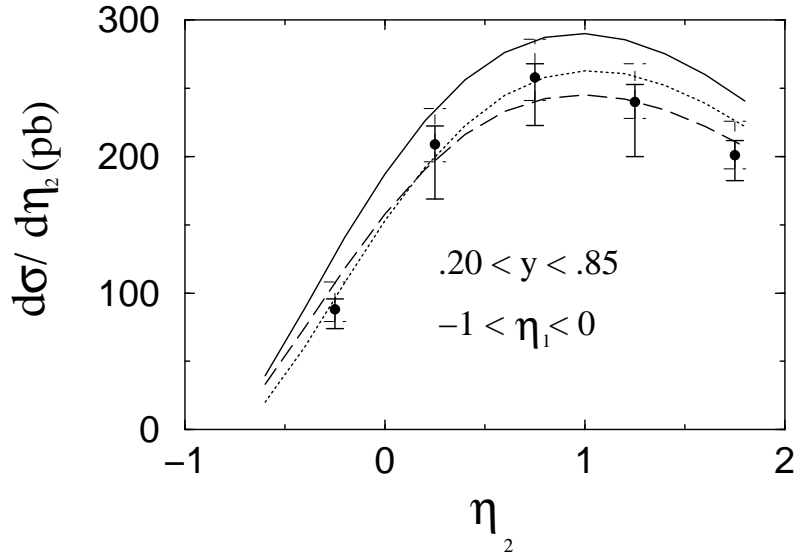}
\end{minipage}
\caption{The cross section $d\sigma/\eta_2$ compared with ZEUS data \cite{16r}. (Full error
bars~: statistical errors~; dashed error bars~: systematic errors including energy-calibration
errors). Full line: $.2 \leq y \leq .85$. Dotted line: $.2 \leq y \leq .80$. The dashed line is obtained by reducing the
distribution $g/P$ by 20~\%.} \label{fig:6} \end{figure}

In Figure \ref{fig:6} we present the results of a similar study done for negative values of
$\eta_1$. In this kinematical range, we can infer from fig.~\ref{fig:6} that the cross
section depends very little on moderate changes of the gluon distribution in the photon.
Therefore this kinematical range is well-suited for a study of the gluon in the proton at
$x_p \sim 2E_T/2E_p \sim 15/880 = .018$. Here we choose data corresponding to the following $y$-range: $.2 < y <
.85$. Also displayed are predictions obtained for $.2 \leq y \leq .8$, and those obtained
with a gluon distribution in the proton multiplied by a factor $.8$. We see that a small
modification of the $y$-range produces an effect similar to a change of the gluon
distribution by 20~\%. Here again, the experimental errors are larger than the effects due
to a modification of the theoretical inputs, but there is a slight indication that data
prefer a smaller gluon distribution in the proton.

Finally we turn to the $E_{T}$-spectrum obtained by the ZEUS collaboration. Here again we
choose to study the large rapidity region with the dijet cross section integrated over the
ranges $1 \leq \eta_1 \leq 2$ and $1 \leq \eta_2 \leq 2$:

\begin{eqnarray} 
d\sigma /dE_{T}^{leading} = \int_1^2 d\eta_1 \int_1^2 d\eta_2 \ {d\sigma
\over dE_{T}^{leading} \ d\eta_1 \ d\eta_2}\label{10e} 
\end{eqnarray}

\noindent $E_{T}^{leading}$ is the transverse energy of the leading jet (highest $E_{T}$).
The transverse energy of the other jet is constrained by the condition 11 GeV $< E_{T} <
E_{T}^{leading}$. Our results are shown in Figure 7. Predictions for the range $.25 \leq y
\leq .85$ and for a gluon in the photon increased by 20~\% are also shown. A slightly better
agreement with data is obtained in the latter case (hardly distinguishable on a logarithmic
plot).

The H1 collaboration presented results \cite{14r} under a form which is very close to the
one advocated in the preceding section~; the authors chose to give the cross section 
$d\sigma/(dx_{\gamma}^{obs} d \ {\rm Log} ((E_{T}^{jets})^2/{\rm GeV}^2))$ with the
definitions
 
\begin{eqnarray}
E_{T}^{jets} = {E_{T 1} + E_{T 2} \over 2}
\label{8e}
\end{eqnarray}

\noindent where $E_{T 1}$ and $E_{T 2}$ are the transverse energies of the two jets with the highest transverse energies in an event. Other kinematical requirements are

\begin{eqnarray}
{|E_{T 1} - E_{T 2}| \over 2 E_{T}^{jets}} &<& .25 \nonumber \\
0 < {\eta_1 + \eta_2 \over 2} &<& 2 \nonumber \\
|\eta_1 - \eta_2| &<& 1 \quad .
\label{9e}
\end{eqnarray}

\noindent The cuts on the ``real'' photon variables are $Q_{max}^2 = 4\ {\rm GeV}^2$
and $.2 < y < .83$. We use a
cone algorithm \cite{15r} with $R = .7$ and no $R_{sep}$ in agreement with ref. \cite{14r}. Moreover we
avoid double counting of jet configurations by choosing the jet of highest $E_{T}$
(made of two partons) when the parton configuration also allows to construct two jets
(made of one parton each). The scales $M$ and $\mu$ are set equal to $E_{T}^{jets}$. \par

\begin{figure}
\centering
\begin{minipage}{0.7\textwidth}
\centering
\includegraphics[width=0.9\textwidth]{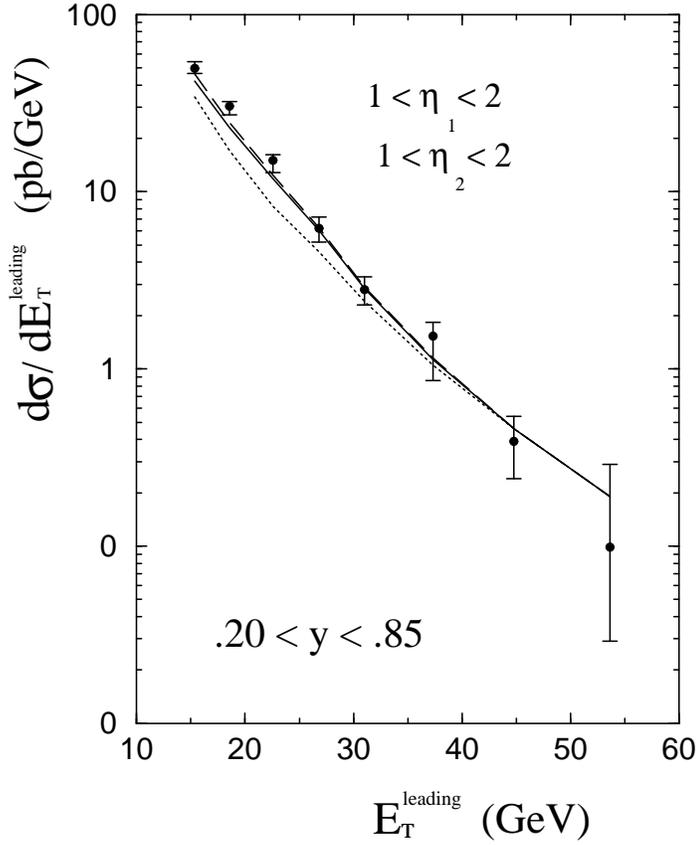}
\end{minipage}
\caption{The cross section $d\sigma/dE_{T}^{leading}$ compared to ZEUS data \cite{16r}. (In
this figure the energy-calibration errors are not included in the error bars). Full line: $.20
\leq y \leq .85$. Dotted line: $.25 \leq y \leq .85$. The dashed line is obtained by increasing
the gluon distribution $g/\gamma$ by 20~\%.} \label{fig:7} \end{figure}

\begin{figure}
\centering
\begin{minipage}{\textwidth}
\centering
\includegraphics[width=0.6\textwidth]{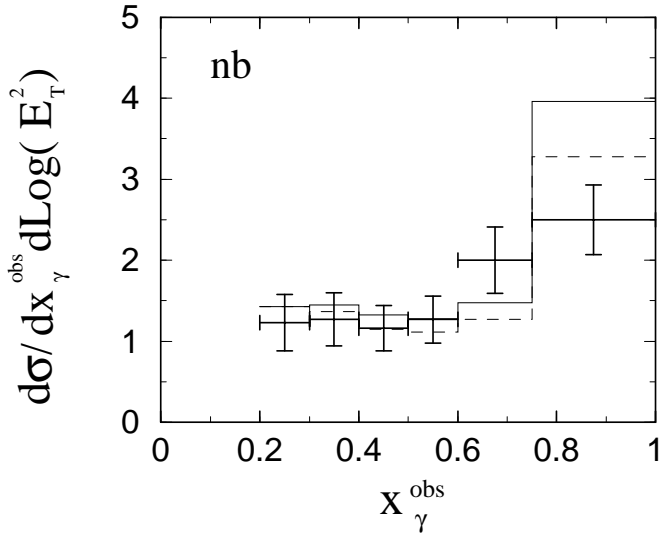}
\end{minipage}
\caption{The cross section $d\sigma/(dx_{\gamma}^{obs} d {\rm Log} ((E_{T}^{jet}/{\rm
GeV})^2))$ in the range $2.3 \leq {\rm Log} (E_{T}^{jet}/{\rm GeV})^2 \leq 2.5$
\protect{\cite{14r}}. Full curve~: $.2 \leq y \leq .83$~; dashed curve~: $.25 \leq y \leq
.83$.} \label{fig:8} \end{figure}

In Fig.~\ref{fig:8} we compare our results with H1 data in the
range $2.30 < \  {\rm Log} \left ( \frac{E_{T}^{jets}}{GeV} \right )^2
\leq 2.50$ where $E_{T}^{jets}$ is large enough to
allow us to neglect (in a first approximation) hadronisation effects and underlying event
contributions. A good agreement\footnote{Unlike other observables, here we disagree with the
predictions of ref. \protect{\cite{21r}}. Our values are higher at small
$x_{\gamma}^{obs}$, by a few tens of percents~; this cannot be explained by the different values of $N_f$ used in the calculations (4 flavors in ref. \protect{\cite{21r}} and 5 in this paper).} is obtained with data, except for $x_{\gamma}^{obs}$ close to one (the
``Direct'' domain) where theory overshoots data. It must be again noted that the theoretical curve
is very sensitive to the photon energy range $y = E_{\gamma}/E_e$. A change of the lower
limit from $y = .20$ to $y = .25$ leads to the dashed line in fig.~\ref{fig:8}~; the cross section
decreases for $x_{\gamma}^{obs} \simeq 1.0$. However it is unlikely that the disagreement for $x_{\gamma}^{ob} \sim 1$ between theory and
H1 data could be explained by a dispersion of the upper and low bounds in $y$. (The
variation of $y_{low}$ by 25~\% studied above is certainly bigger than the experimental
systematic error). A decrease of the gluon distribution in the proton would decrease
$d\sigma/dx_{\gamma}^{obs}$ in all bins in $x^{obs}_{\gamma}$. It is compatible with data at
low $x_{\gamma}^{obs}$, but its effect at large $x_{\gamma}^{obs}$ is not sufficient to put
theory in agreement with data. The resolved contribution is also important in the highest
$x_{\gamma}^{obs}$-bin, a kinematical region which explores the quark contents of the photon
at large $x_{\gamma}$. In this domain the quark distribution is given by the pointlike
component (non pointlike contributions of the ``Vector Meson Dominance'' type are negligible)
which cannot be modified in an arbitrary way, and adjusted to data. For instance the
difference between the AFG and GRV parametrization is small in this region (see
fig.~3). \par

The first conclusion that we can draw from this study of H1 and ZEUS data is that overall
there is a good agreement between experiment and theory. However the systematic errors are
non-negligible and correspond roughly to variations of the theoretical predictions coming
from modifications by $\pm$ 20~\% of the gluon distribution normalizations. Therefore it appears
difficult to constrain the gluon distributions with an accuracy better than a few tens of
per cents with the present data. 

Part of the errors should cancel in the ratio

\begin{eqnarray}
dN/dx_{\gamma}^{obs} = {d\sigma /dx_{\gamma}^{obs} \over \int_0^1 dx_{\gamma}^{obs} \ d\sigma /dx_{\gamma}^{obs}}
\label{11e}
\end{eqnarray}

\noindent which could allow a better determination of the $x$ shape of the distributions functions. However we would thus loose any information on the absolute normalization.
 
\section{Conclusions}
\label{conclusion} In this paper we described a new NLL event generator for photoproduction
reactions involving the direct and resolved contributions, and we used it to assess the
possibility to measure the quark and gluon contents of the photon from photoproduction
experiments. (Actually we assumed that the quark distributions in the photon were fixed from
$\gamma\gamma^*$ DIS experiments and we concentrated on the gluon contents).
 \par

The cross section $d\sigma/d\bar{x}_{\gamma}$, where $\bar{x}_{\gamma}$ is related to the
scaled momentum $x_{\gamma}$ of partons in the photon, is quite appropriate to a study of these
contents. We discussed two definitions of $\bar{x}_{\gamma}$~: $\bar{x}_{\gamma} =
x_{\gamma}^{obs}$, the well-known definition of the ZEUS collaboration, and $\bar{x}_{\gamma} =
x_{\gamma}^{LL}$, a variable which reduces problems of infrared sensitivity. Then we showed
that the cross section $d\sigma /dx_{\gamma}^{obs}$ is sensitive to the gluon
density of the photon only if we require the two jets with the highest transverse
energy to have positive rapidities. \par

A relatively good agreement between theory and experiment is found when confronting the predictions,
obtained with the CTEQ4M and GRV distributions, with H1 and ZEUS data. It is interesting to
note that H1 and ZEUS data are compatible with a 20~\% increase of the gluon contents of the
photon (Fig.~6, $\eta_1 \sim 1.5$, Fig.~8, $E_{T}^{leading} \sim  20$~GeV) and with a 20~\%
decrease of the gluon contents of the proton (Fig.~5, $x_{\gamma}^{obs}
\sim 1$ where the direct contribution is important~; Fig.~7).
Unfortunately this remark cannot be made more quantitative, because the systematic errors,
essentially coming from the uncertainties in the measurement of the jet energies, are large and of the same
order as the variations of the theoretical predictions due to gluon distribution
modifications. One must also keep in mind the great sensitivity of the cross
section to the photon energy-range. The knowledge of the resolution of the
variable $y$ (obtained from the Jacquet-Blondel variable $y_{JB}$) should allow a
more accurate prediction. However it is interesting to remark that 1996 and 1997
preliminary ZEUS data \cite{17r} also favor a larger gluon distribution in the
photon. \par

Until now we have not discussed the theoretical uncertainties coming from the scale dependence
of the cross sections. These uncertainties were carefully studied in ref. \cite{9bisr} and in ref.
\cite{21r}, and found to be of the order of a few tens of percents for large variations of the
values of the factorization and renormalization scales. In this paper we continue this study by
looking at the sensitivity of the cross section $d\sigma/(dx_{\gamma}^{obs} d {\rm Log}
(E_{T}^{jets})^2)$ which can be compared with H1 data (Fig.~5)~; we make this study for the
range $.75 \leq x_{\gamma}^{obs} \leq 1$ and with the kinematical conditions of Fig.~5. The
scales are $M = \mu = \kappa E_{T}^{jets}$ and $\kappa$ is varied between $.5$ and $2.$ Our
results are summarized in table 1.

\begin{center}
\begin{tabular}{|c|c|c|c|}
\hline
$\kappa$ &{\bf Direct contribution} &{\bf Resolved contribution} &{\bf Total contribution} \\
\hline
.5 &.634 &.384 &1.018 \\
.75 &.578 &.432 &1.010 \\
1.00 &.547 &.453 &1.000 \\
1.50 &.512 &.474 &.986 \\
2.00 &.491 &.482 &.973 \\
\hline
\end{tabular}
\end{center}

\noindent {\bf Table 1} : The sensitivity of the direct and resolved photoproduction cross-sections
to the renormalization and factorization scale $M = \mu = \kappa \cdot E_{T}^{jets}$ (normalized to the total contribution at $\kappa = 1.0$). \\

We note that the cross section $d\sigma/dx_{\gamma}^{obs}$ is quite stable and varies by less than
5~\% when the scales $M^2 = \mu^2$ vary by a factor 16. The theoretical errors appear to be well
under control, at least for this observable. \par

So we can conclude that, in the future, the possibility to accurately determine the gluon distribution
in the photon relies on the possibility to improve the experimental systematic errors. In the
future also statistics will be larger and higher $E_{T}$ regions will be accessible. The
corresponding data will put more constraints on the parton distributions. \\

\noindent {\bf Acknowledgments} \par
 We would like to thank Joost Vossebeld for an interesting correspondence on ZEUS
results.


\begin{thebibliography}{bibli}
\bibitem{AFG-1jet} P. Aurenche, M. Fontannaz and J. Ph. Guillet, Phys. Lett. {\bf B338} (1994) 98. It should be noticed that the resolved contribution  shown in this paper is too high by a factor 2. This has been corrected for the  actual comparison.
\end{thebibliography}
\end{document}